\documentclass[fleqn,usenatbib,useAMS]{mnras}
\usepackage{graphicx}   
\usepackage{amsmath}    
\usepackage{amssymb}    
\usepackage{multicol}        
\usepackage{bm}         
\usepackage{pdflscape}  
\usepackage[T1]{fontenc}
\usepackage{ae,aecompl}
\usepackage{newtxtext,newtxmath}

  \def\aap{A\&A} \def\apjl{ApJ}
\def\apj{ApJ} \def\apjs{ApJS} \def\aj{AJ} \def\mnras{MNRAS} 
\def\pasp{PASP} \def\apss{Ap\&SS}

\title[{\it GALEX} Absolute Calibration] {{\it GALEX} Absolute Calibration
and Extinction Coefficients Based on White Dwarfs}

\author[R. Wall et al.] 
{Renae E. Wall$^{1}$\thanks{\tt rnenwall@ou.edu}, 
Mukremin Kilic$^{1}$, 
P. Bergeron$^{2}$,
B. Rolland$^{2}$,
\newauthor C. Genest-Beaulieu$^{2}$,
A. Gianninas$^{3}$\\ 
$^{1}$Homer L. Dodge Department of Physics \& Astronomy, University of Oklahoma,
440 W. Brooks St, Norman, OK 73019, USA\\
$^{2}$D$\acute{e}$partement de Physique, Universit$\acute{e}$ de Montr$\acute{e}$al, C. P. 6128, Succ. Centre-Ville, Montr$\acute{e}$al, QC H3C 3J7, Canada\\ 
$^{3}$Department of Physics and Astronomy, Amherst College, 25 East Drive, Amherst, MA 01002\\}

\date{\ \ Submitted \today \vspace{-0.5cm}}
\pubyear{2019}

\begin{document}
\label{firstpage}
\pagerange{\pageref{firstpage}--\pageref{lastpage}}
\maketitle

\begin{abstract}

We use 1837 DA white dwarfs with high signal to noise ratio spectra and {Gaia} parallaxes to verify the absolute
calibration and extinction coefficients for the Galaxy Evolution Explorer
({\em GALEX}). We use white dwarfs within 100 pc to verify the 
linearity correction to the {\em GALEX} data. We find that the linearity correction is valid for magnitudes brighter than 15.95  
and 16.95 for the Far Ultraviolet (FUV)
and Near Ultraviolet (NUV) bands, respectively. We also use DA white dwarfs beyond 250
pc to calculate extinction coefficients in the FUV and NUV bands; 
$R_{\rm FUV}=8.01 \pm 0.07$ and $R_{\rm NUV}=6.72 \pm 0.04$.
These are consistent with the predicted extinction coefficients for Milky Way type dust in the FUV, but smaller than predictions in the NUV.
With well understood optical spectra and state-of-the-art model atmosphere analysis,
these white dwarfs currently provide the best constraints on the extinction coefficients
for the {\em GALEX} data. 

\end{abstract}


\section{Introduction}

The {\em Galaxy Evolution Explorer} ({\em GALEX}) is the first space based mission to
attempt an all-sky imaging survey in the ultraviolet \citep[UV,][]{Martin2005}.
In the ten years that it was
operational, {GALEX} surveyed 26,000 square degrees of the sky as part of the
All-sky Imaging Survey in two
band passes: Far Ultraviolet (FUV) with a central wavelength of 1528 \AA\ and
Near Ultraviolet (NUV) with a central wavelength of 2271 \AA\
\citep{Morrissey2005}. Although its primary goal was to study star formation and
galaxy evolution, the first science goal was to determine UV calibration,
particularly extinction \citep{Martin2005}.

There are two sources of nonlinearity in {\em GALEX} photometry: global nonlinearity due to
the finite period required for the electronics to assemble photon lists and
local nonlinearity near bright sources. \citet[][see their Fig. 8]{Morrissey2007}
demonstrate that nonlinearity becomes significant ($>10$\%) above
109 and 311 counts s$^{-1}$ in the FUV and NUV bands, respectively. These correspond
to $m_{\rm FUV} \approx 14$ mag and $m_{\rm NUV} \approx 15$ mag. While the first nonlinearity
is well understood, the second (local nonlinearity) complicates the standard
star measurements. 

{\em GALEX} observed 18 white dwarfs from the {\em Hubble Space
Telescope} CALSPEC database \citep{Bohlin2001} as standard stars. However,
its photometric calibration relies primarily on the dimmest star in this sample,
LDS 749b, as all of the other standard stars observed are highly saturated.
In fact, after \citet{Bohlin2008} provided a better CALSPEC spectrum for LDS 749b,
the {\em GALEX}
magnitudes were shifted by $\approx$0.04 mag between the GR4/5 and GR6 data releases.
Hence, it
is important to verify the photometric calibration using fainter stars.

\citet{Holberg2014} verified the {GALEX} photometric calibration using 
99 and  107 DA white dwarfs in the FUV and NUV, respectively, with magnitudes
between 10 and 17.5 from the final GR7 {GALEX} data release.
They found that a modest linearity correction is needed in this magnitude range. 
Although \citet{Holberg2014} postulate
that their linearity correction should hold for stars as faint as 20th
magnitude, they point out the need for a larger sample size and the
characterization of extinction in the {GALEX} bands. In this work, we
investigate the validity of the \citet{Holberg2014} linearity correction for a
large sample of DA white dwarfs from the Sloan Digital Sky Survey (SDSS), particularly 
probing the fainter magnitudes.

There is a broad range of {\em GALEX} extinction coefficients reported in
the literature. These coefficients are defined as
\begin{equation} 
R_\lambda=\frac{A_\lambda}{E(B-V)},
\label{eq:R}
\end{equation} 
where $A_\lambda$ is the total absorption along the line of sight to an object and $E(B-V)$
is the reddening. \citet{Bianchi2011} provided theoretical estimates of
$R_{\rm FUV} = 8.06$ and $R_{\rm NUV} = 7.95$ for Milky Way type dust and $R_{\rm FUV} = 12.68$
and $R_{\rm NUV} = 8.08$ for the Small Magellanic Cloud (SMC) type dust. 
\citet{Yuan2013} found empirical values of $R_{\rm FUV} = 4.37-4.89$ and
$R_{\rm NUV} = 7.06-7.24$. The latter authors used
the `standard pair' technique \citep{Stecher1965}, where two stars of the
same spectral type, one in an area with low extinction and one in an area of high
extinction, are compared. Those stars with low extinction, the control sample,
are used to determine the intrinsic colors of the corresponding stars with high
extinction, the target sample. \citet{Yuan2013} examined a target sample of 1396
stars and a control sample of 16405 stars from the {GALEX} fifth data
release. Most of these stars were classified as FGK dwarfs, with a small fraction
of A dwarfs and KM giants. However, there is a great deal of scatter and uncertainty
in their derivation of $R_{\rm FUV}$ and $R_{\rm NUV}$, and \citet{Yuan2013} caution
against using their {GALEX} extinction coefficients.
In this work, we re-derive the
{GALEX} extinction coefficients using a large sample of DA white dwraf stars
with high S/N SDSS spectra and {Gaia} parallaxes to obtain a more reliable estimate.

We present the white dwarf sample used in this study in Section \ref{sec:data}
and describe our calculation of the synthetic magnitudes in Section \ref{sec:synth}.
Our analysis of nonlinearity is presented in Section \ref{sec:lin}
followed by our derivation of the \textit{GALEX} extinction coefficients in Section \ref{sec:R}.
The 3$\sigma$ outliers are discussed in Section \ref{sec:out}, and we conclude in Section \ref{sec:con}.

\section{Sample Selection} \label{sec:data}

In order to improve calibrations for the \textit{GALEX} data, we select all
spectroscopically confirmed DA white dwarfs from the SDSS data releases 7, 10,
and 12 with S/N $\ge 20$ spectra
\citep{Kleinman2013,Kepler2015,Kepler2016}. This selection insures that the
$T_{\rm eff}$ and $\log{g}$ measurements are precise enough to model
the emergent stellar fluxes in the UV bands. We focus on DA white dwarfs due to
our good understanding of their opacities and atmospheres \citep{Holberg2006}. 
We cross reference our initial sample of 3733 DA white dwarfs from the SDSS with Pan-STARRS, and
we cross reference our sample once more with \textit{Gaia DR}2, selecting all stars 
with parallax/error $\ge 5$. We then cross reference our sample with
the \textit{GALEX} catalog of unique UV sources from the 
All Sky Imaging Survey (GUVcat) presented in \citet{Bianchi2017}.
We use a search radius of $2\arcsec$ and find a total of 1837 stars with \textit{GALEX}
photometry.

We break our initial sample of stars into two groups based on \textit{Gaia} distance: stars
within 100 pc and stars further than 250 pc. There are 339 (627) and 451 (628) stars with FUV and NUV photometry in the
100 ($d>250$) pc sample, respectively. We leave the examination of stars between
100 and 250 pc for future work. The local interstellar medium is
relatively devoid of cold neutral gas, up to about 100 pc, the boundary of the
Local Bubble \citep{Lallement2003,Redfield2006}. Since extinction is not an
issue for the 100 pc sample, we use it to verify the {\em GALEX} photometric
calibration. The $d>250$ pc sample suffers from full extinction, and we use it
to calculate the extinction coefficients in both the FUV and NUV bands.

\section{Synthetic Magnitudes}\label{sec:synth}

\citet{Genest-Beaulieu2019} found a systematic offset between temperatures derived
using the spectroscopic \citep{Bergeron1992} and photometric \citep{Bergeron1997} techniques.
They determine that this offset is caused by inaccuracies in the treatment
of Stark broadening in their model spectra. The photometric technique is
less sensitive to the input physics of the models, so we adopted it for this work.
We use SDSS \textit{u} and Pan-STARRS {grizy} photometry and \textit{Gaia} parallaxes to derive photometric
temperatures and radii for all stars in our final sample. 
These temperature and radius measurements are then used to calculate a model
spectrum for each white dwarf in the 100 pc sample.

To estimate the average flux
in a given bandpass, $f^{m}_{\lambda}$, we use the equation

\begin{equation}
f^{m}_{\lambda} = \frac{\int^{\infty}_{0}f_{\lambda}S_{m}(\lambda)\lambda {\rm d}\lambda}{\int^{\infty}_{0}S_{m}(\lambda)\lambda {\rm d}\lambda},
\end{equation}

\noindent where $S_{m}(\lambda)$ is the transmission function of the
corresponding bandpass, and $f_{\lambda}$ is the monochromatic flux from
the star received at Earth \citep{Bergeron1997,Gianninas2011}.
SDSS, Pan-STARRS, and {\em GALEX} use the AB magnitude system. We transform the
average flux in a given bandpass to an average magnitude using
the equation

\begin{equation}
m = -2.5\log f^{m}_{\nu} - 48.6. 
\end{equation}

This procedure enables us to calculate the absolute magnitude of each star
in each filter. We use
the observed and dereddened SDSS magnitudes for the $d<100$ and $d>250$ pc samples,
respectively.

\section{{\em GALEX} Photometric Calibration}
\label{sec:lin} 

\begin{figure}
\includegraphics[width=\columnwidth, bb=38 5 659 539]{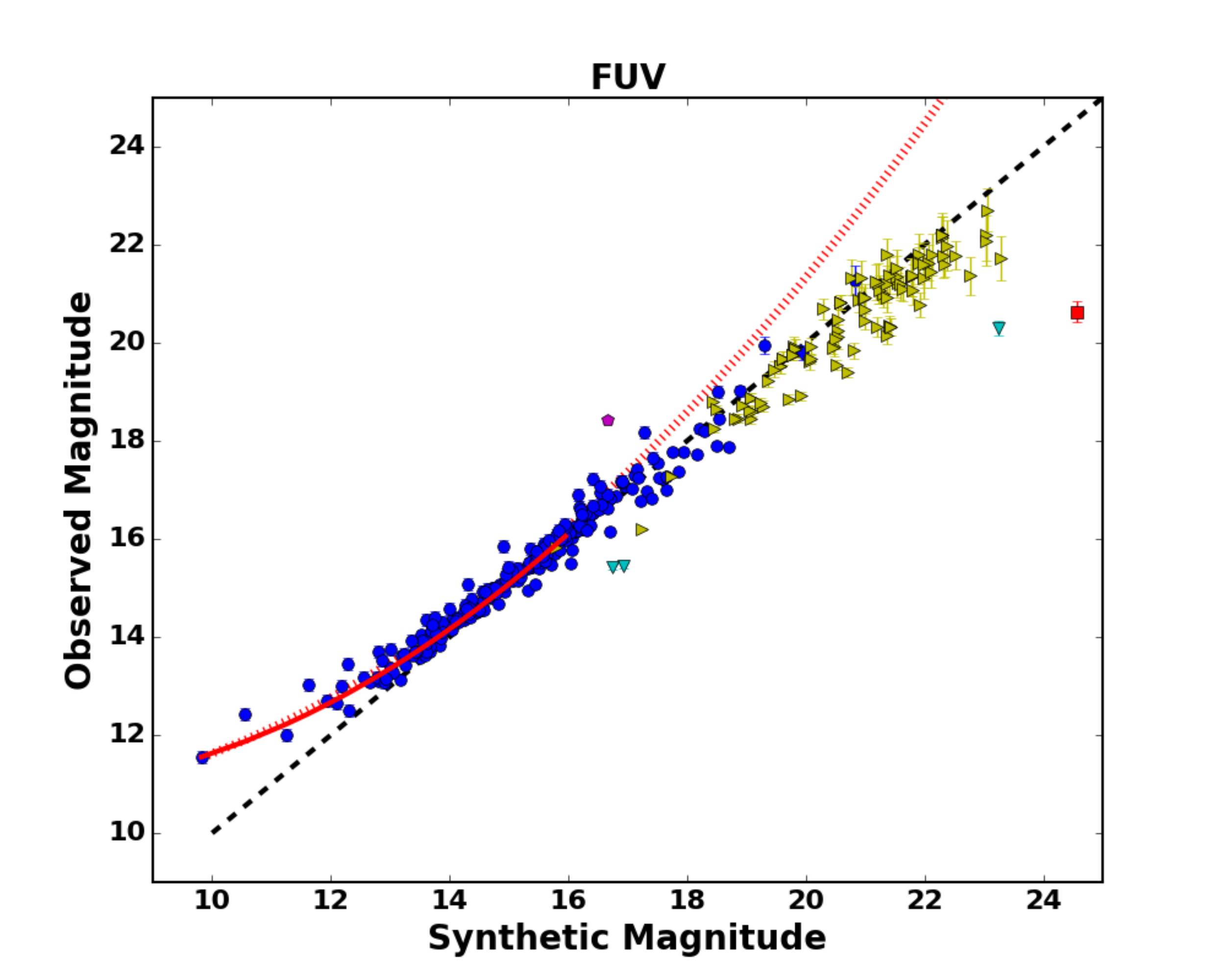}
\includegraphics[width=\columnwidth, bb=38 5 659 539]{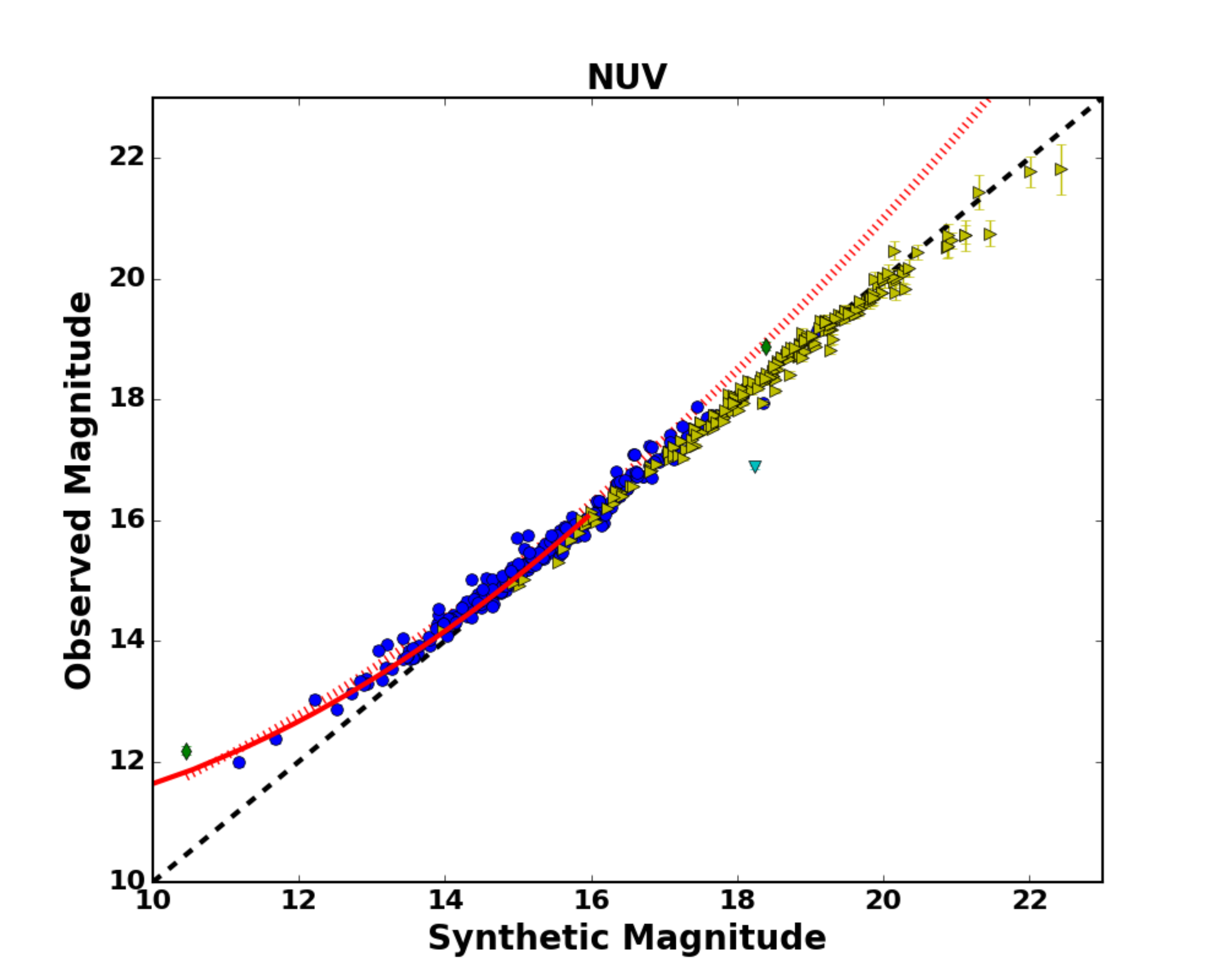}
\caption{The linearity fit for stars within 100 pc in the FUV
band (top) and the NUV band (bottom). The quadratic linearity fit from
\citet{Holberg2014} is marked as the red dashed line. Our linearity fit is
plotted in solid red. The black dashed line is the one-to-one correlation.
Stars with $T_{\rm eff}$ below 11,000 K are plotted as yellow triangles.
Cyan triangles are known double degenerate systems and green diamonds are
previously known WD+main sequence binaries. The magenta pentagon marks
the ZZ Ceti WD1258+013. Previously unknown 3$\sigma$ outliers are plotted as red squares.}
\label{fig:lin}
\end{figure}

\begin{table*}
        \centering \caption{Properties of stars within 100 pc in our sample. Stars with WD names are from \citet{Gianninas2011}. The full table is available online.} \label{tab:100sample}
        \begin{tabular}{lccccc}
                \hline Star & Synthetic FUV mag & Synthetic NUV mag & Observed FUV mag & Observed NUV mag\\ \hline
		J000410.42-034008.60 & 25.802 & 19.201 & \dots & 19.142 $\pm$ 0.070\\
		J001339.11+001924.90 & 19.769 & 16.403 & 19.756 $\pm$ 0.103 & 16.449 $\pm$ 0.014\\
		J002049.39+004435.10 & 22.29 & 17.96 & 22.194 $\pm$ 0.369 & 17.937 $\pm$ 0.027\\
		WD0023-109 & 20.714 & 17.385 & 19.394 $\pm$ 0.093 & 17.203 $\pm$ 0.019\\
		J002634.39+353337.60 & 23.59 & 19.187 & \dots & 19.132 $\pm$ 0.072\\
		J003328.03+054039.18 & 26.94 & 20.354 & \dots & 20.167 $\pm$ 0.142\\
		J003511.63+001150.40 & 20.592 & 17.417 & 20.817 $\pm$ 0.151 & 17.511 $\pm$ 0.022\\
		WD0033+016 & 18.921 & 16.312 & 18.727 $\pm$ 0.077 & 16.368 $\pm$ 0.016\\
		WD0037-006 & 16.92 & 15.548 & 15.456 $\pm$ 0.016 & 15.299 $\pm$ 0.010\\
		J004511.19+090445.37 & 25.926 & 19.923 & \dots & 19.922 $\pm$ 0.128\\
		WD0048+202 & 14.519 & 14.839 & 14.565 $\pm$ 0.008 & 14.837 $\pm$ 0.005\\
		J005438.84-095219.70 & 22.299 & 18.077 & 22.158 $\pm$ 0.497 & 17.942 $\pm$ 0.043\\
		WD0058-044 & 14.933 & 15.139 & 14.934 $\pm$ 0.016 & 15.182 $\pm$ 0.011\\
		WD0100-036 & 22.062 & 17.91 & 21.635 $\pm$ 0.369 & 17.885 $\pm$ 0.039\\
		WD0101+059 & 16.318 & 16.138 & 16.219 $\pm$ 0.021 & 16.132 $\pm$ 0.012\\
		WD0101+048 & 20.513 & 15.622 & 19.545 $\pm$ 0.103 & 15.528 $\pm$ 0.011\\
		WD0102+095 & 13.163 & 13.626 & 13.130 $\pm$ 0.007 & 13.801 $\pm$ 0.006\\
		J010543.14-092054.60 & 24.111 & 18.849 & \dots & 18.729 $\pm$ 0.046\\
		WD0104+015 & 23.286 & 18.723 & 21.735 $\pm$ 0.448 & 18.408 $\pm$ 0.063\\
		WD0107+267 & 15.156 & 15.181 & 15.144 $\pm$ 0.008 & 15.231 $\pm$ 0.005\\ \hline
        \end{tabular}
	
\end{table*}

Our 100 pc SDSS sample contains few stars brighter than 14th magnitude. In order
to constrain the fit for both faint and bright white dwarfs, we extend our 
sample to include the 100 pc white dwarfs from \citet{Holberg2014} and \citet{Gianninas2011}. 
Figure \ref{fig:lin} compares the observed and predicted synthetic magnitudes
for this sample in
both bands. The solid and dashed lines show a quadratic polynomial fit to the data
and the one-to-one line, respectively. Stars with $T_{\rm eff}$ below 11,000 K
are represented by yellow triangles.  Stars below this temperature suffer from the
red wing of the Ly$\alpha$ opacity,
which affects the ultraviolet more strongly than the optical \citep{Kowalski2006}. 
The 3$\sigma$ outliers that are known double degenerates, white dwarf + main sequence
binaries, and ZZ Cetis are marked by cyan triangles, green diamonds, and
magenta pentagons,respectively. Previously unknown 3$\sigma$ outliers in 
our polynomial fit are plotted as red squares. All 3$\sigma$ outliers are excluded from this
fit. We further discuss these outliers in Section \ref{sec:out}. 
  
Our quadratic fits are represented by the expression

\begin{equation}
m_{\rm obs} = c_2 m^{2}_{\rm synth} + c_1 m_{\rm synth} + c_0,
\label{eq:fit}
\end{equation}

\noindent where $m_{\rm obs}$ and $m_{\rm synth}$ are the observed and synthetic 
\textit{GALEX} magnitudes, respectively. The best fit values of the fitting coefficients
$c_0$, $c_1$, and $c_2$ are given in Table \ref{tab:synthvsobs}.

\begin{table}
        \centering \caption{Fitting parameters for the linearity correction in the FUV and NUV bands (see Eq \ref{eq:fit}.)} \label{tab:synthvsobs}
        \begin{tabular}{ccc}
                \hline Property & FUV & NUV \\ \hline 
                $c_0$ & 13.23 & 10.49\\
                $c_1$ & -0.727 & -0.31\\
                $c_2$ & 0.057 & 0.041\\ 
                Range & $\leq 15.95$ mag & $\leq 16.95$ mag\\ \hline
        \end{tabular}
\end{table}

\citet{Holberg2014} found a non-linear correlation and small offset between 
{\em GALEX} fluxes and predicted fluxes for their sample. Their quadratic
fit is shown as a dotted line in Figure \ref{fig:lin} and is based on about
100 DA white dwarfs with FUV and NUV magnitudes between
 10 and 17.5 mag. However, they only have 6-8 stars fainter than 17th
magnitude in their sample, hence the fit is relatively unconstrained at the
faint end. The dotted line significantly underpredicts the observed magnitudes
in both FUV and NUV bands, and is clearly not useful below 17th magnitude.

With a significantly larger number of fainter DA white dwarfs, we are able
to test for non-linearities in the data down to magnitudes fainter than 20.
We note that stars with $T_{\rm eff}$ below 11,000 K have systematically 
fainter synthetic magnitudes in the FUV, while there is no systematic offset
in the NUV. Since these stars are affected by the red wing of the Ly$\alpha$ opacity \citep{Kowalski2006},
our results indicate that this opacity source is well handled
in our models for the NUV, while the modeling 
of this opacity should be revisited for the FUV. To remove this systematic effect from our fit,
we first fit the full sample to calculate the magnitude where the full quadratic 
fit crosses the one-to-one line for the FUV and NUV. Only stars brighter than these 
magnitudes, 15.95 mag (FUV) and 16.95 mag (NUV), will require a linearity correction.
To determine the linearity correction, we then fit only stars brighter than
15.95 mag (FUV) and 16.95 mag (NUV). This is our final quadratic fit which is plotted
in Figure \ref{fig:lin}. Our linearity corrections are not statistically different
from those presented in \citet{Holberg2014}. To convert the observed \textit{GALEX} 
magnitudes into corrected magnitudes, we find the quadratic solutions to the
linearity corrections shown in Figure \ref{fig:lin}. Our final corrections take the form

\begin{equation}
m_{\rm corr} = c_0 + (c_1 m_{\rm obs} + c_2)^{1/2},
\label{eq:corr}
\end{equation}

\noindent where $m_{\rm obs}$ and $m_{\rm corr}$ are the observed and corrected 
\textit{GALEX} magnitudes, respectively. The calculated constants
$c_0$, $c_1$, and $c_2$ are given in Table \ref{tab:corr}. These 
corrections are applicable to objects brighter than 15.95 mag and 16.95 mag in the 
FUV and NUV, respectively.

\begin{table}
        \centering \caption{Inverse quadratic corrections for the FUV and NUV bands (see Eq \ref{eq:corr}.)} \label{tab:corr}
        \begin{tabular}{ccc}
                \hline Property & FUV & NUV \\ \hline 
                $c_0$ & 6.412 & 3.778\\
                $c_1$ & 17.63 & 24.337\\
                $c_2$ & -192.135 & -241.018\\ \hline
        \end{tabular}
\end{table}

\section{Extinction Coefficients} 
\label{sec:R} 

\begin{table*}
        \centering \caption{Properties of stars beyond 250 pc in our sample. Stars with WD names are from \citet{Gianninas2011}. The full table is available online.} \label{tab:250sample}
        \begin{tabular}{lccccc}
                \hline Star & Synthetic FUV mag & Synthetic NUV mag & Observed FUV mag & Observed NUV mag\\ \hline
		J000302.59+240555.80 & 18.699 & 18.73 & 19.461 $\pm$ 0.077 & 19.241 $\pm$ 0.039\\
		J000626.69+242441.70 & 18.196 & 18.372 & 18.964 $\pm$ 0.067 & 18.995 $\pm$ 0.031\\
		J001043.55+253829.18 & 15.968 & 16.526 & 16.395 $\pm$ 0.020 & 16.890 $\pm$ 0.010\\
		J001549.44+245604.91 & 15.691 & 16.253 & 16.038 $\pm$ 0.016 & 16.558 $\pm$ 0.009\\
		J001712.70+250443.04 & 17.95 & 18.125 & 18.218 $\pm$ 0.047 & 18.391 $\pm$ 0.026\\
		J002126.69-093714.20 & 17.688 & 17.946 & 18.049 $\pm$ 0.035 & 18.252 $\pm$ 0.024\\
		WD0019+150 & 15.3 & 15.863 & 15.697 $\pm$ 0.018 & 16.160 $\pm$ 0.013\\
		J002636.48-100330.50 & 17.116 & 17.444 & 17.552 $\pm$ 0.050 & 17.757 $\pm$ 0.024\\
		J002806.49+010112.20 & 16.185 & 16.67 & 16.477 $\pm$ 0.028 & 16.885 $\pm$ 0.022\\
		J003533.74+240253.17 & 17.293 & 17.688 & 17.683 $\pm$ 0.057 & 18.139 $\pm$ 0.048\\
		J004346.36+254910.50 & 18.482 & 18.482 & 18.695 $\pm$ 0.120 & 18.647 $\pm$ 0.065\\
		J004648.66+250915.10 & 17.924 & 18.095 & 18.405 $\pm$ 0.091 & 18.525 $\pm$ 0.043\\
		J005547.78-084507.30 & 13.795 & 14.376 & 14.885 $\pm$ 0.014 & 15.160 $\pm$ 0.011\\
		J010810.17+183120.46 & 16.957 & 17.351 & 17.586 $\pm$ 0.065 & 17.956 $\pm$ 0.044\\
		J011100.64+001807.20 & 19.873 & 19.182 & 20.996 $\pm$ 0.320 & 19.425 $\pm$ 0.109\\
		J011428.32+215310.79 & 16.21 & 16.764 & 16.439 $\pm$ 0.034 & 17.012 $\pm$ 0.027\\
		J011541.62+310404.20 & 16.432 & 16.952 & 16.963 $\pm$ 0.042 & 17.318 $\pm$ 0.030\\
		J012041.19+395307.20 & 17.423 & 17.683 & 17.671 $\pm$ 0.067 & 17.923 $\pm$ 0.033\\
		J012318.14+330014.34 & 17.713 & 17.892 & 17.927 $\pm$ 0.043 & 18.12 $\pm$ 0.029\\
		J012601.53+332523.49 & 16.21 & 16.702 & 16.917 $\pm$ 0.042 & 17.252 $\pm$ 0.022\\ \hline
        \end{tabular}
	
\end{table*}

\begin{figure}
\includegraphics[width=\columnwidth, bb=38 5 659 539]{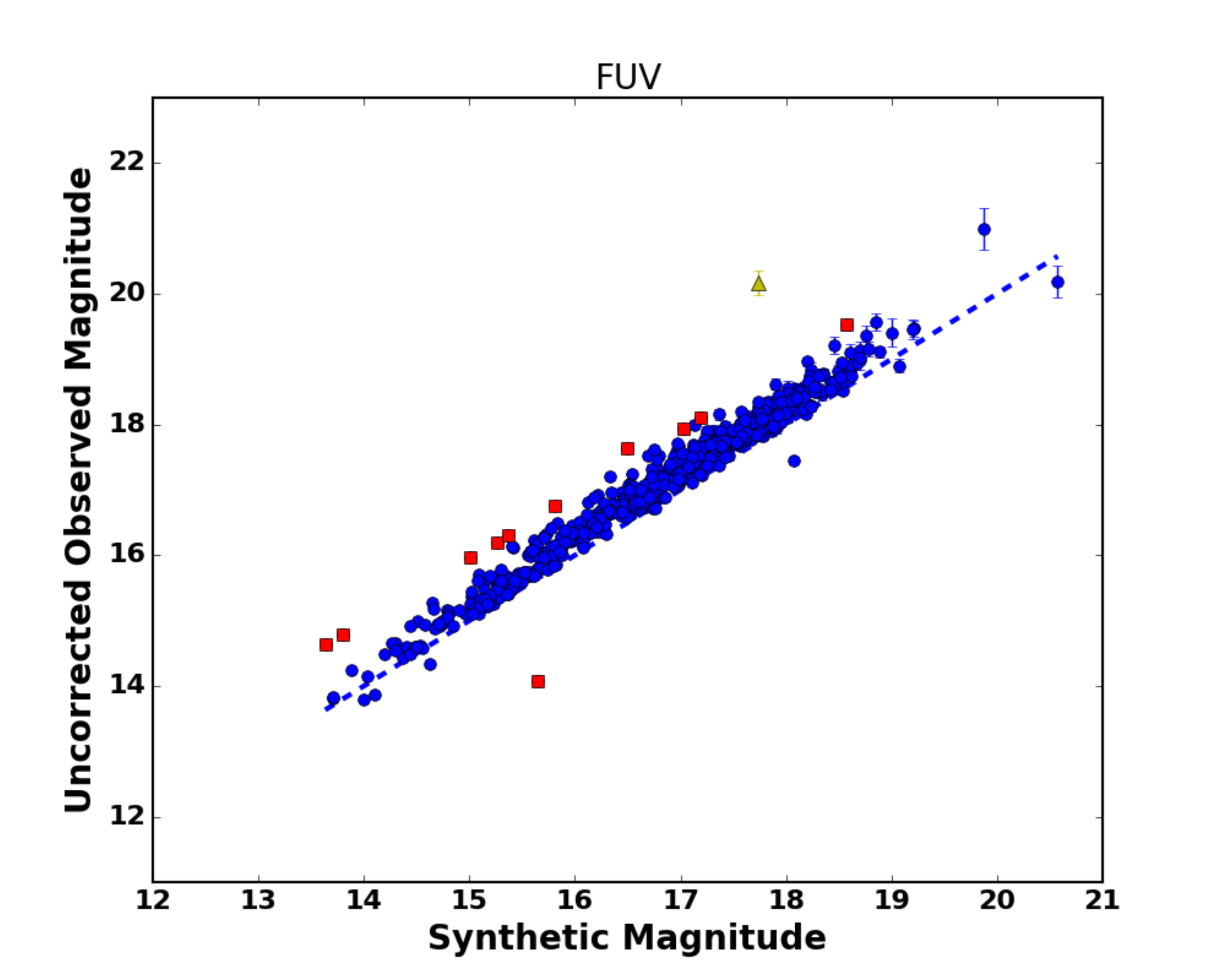}
\includegraphics[width=\columnwidth, bb=38 5 659 539]{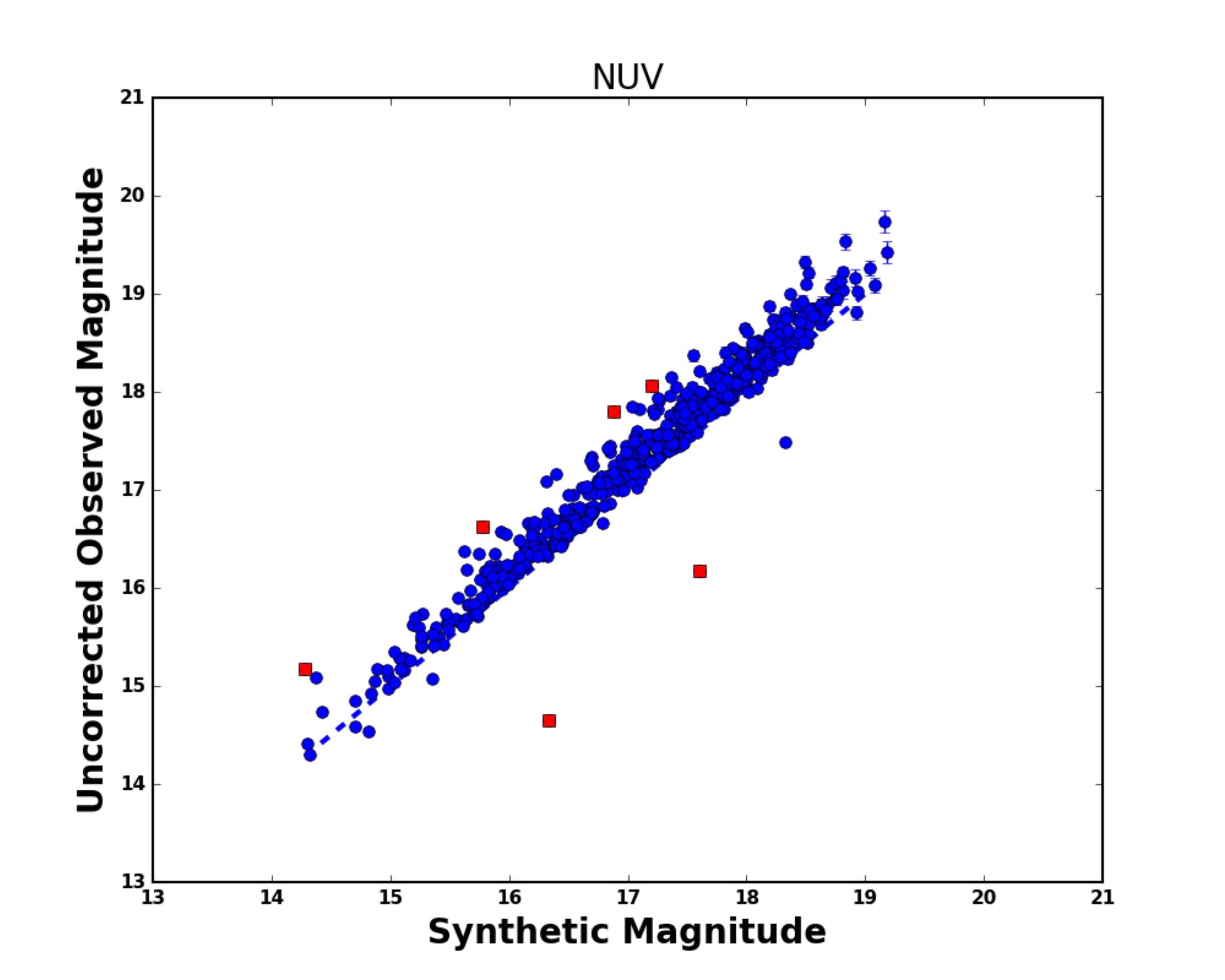}
\caption{Uncorrected versus model magnitudes for stars beyond
250 pc in the FUV and NUV bands. The blue dashed
line is the one-to-one correlation and 4$\sigma$ outliers are plotted
as red squares. The candidate double degenerate J211607.27+004503.17 is plotted as a yellow triangle.}
\label{fig:without}
\end{figure}

After revisiting the linearity corrections and determining the magnitudes they are valid
over, we examine the sample of SDSS DA white dwarfs with {Gaia} distances beyond  250 pc. We apply
the linearity corrections given in Table \ref{tab:corr} only to those
stars brighter than our cut-off magnitudes. Figure \ref{fig:without} shows the observed
versus synthetic magnitudes for the 250 pc white dwarf sample.
These stars experience full extinction, which leads to observed FUV and NUV
photometry fainter than expected. 

We calculate the $R$ value in the
NUV and FUV bands for each star using Equation \ref{eq:R}, the total absorption
in each filter $A_\lambda$ (the difference between the synthetic and observed magnitude), 
and $E(B-V)$ from \citet{Schlafly2011}. We find 9 stars
in the FUV and 18 stars in the NUV with negative $R$ values. These stars, as well as
the 4$\sigma$ outliers, are excluded 
from the weighted average of the $R$ vaules.

\begin{figure}
\includegraphics[width=\columnwidth, bb=38 5 659 539]{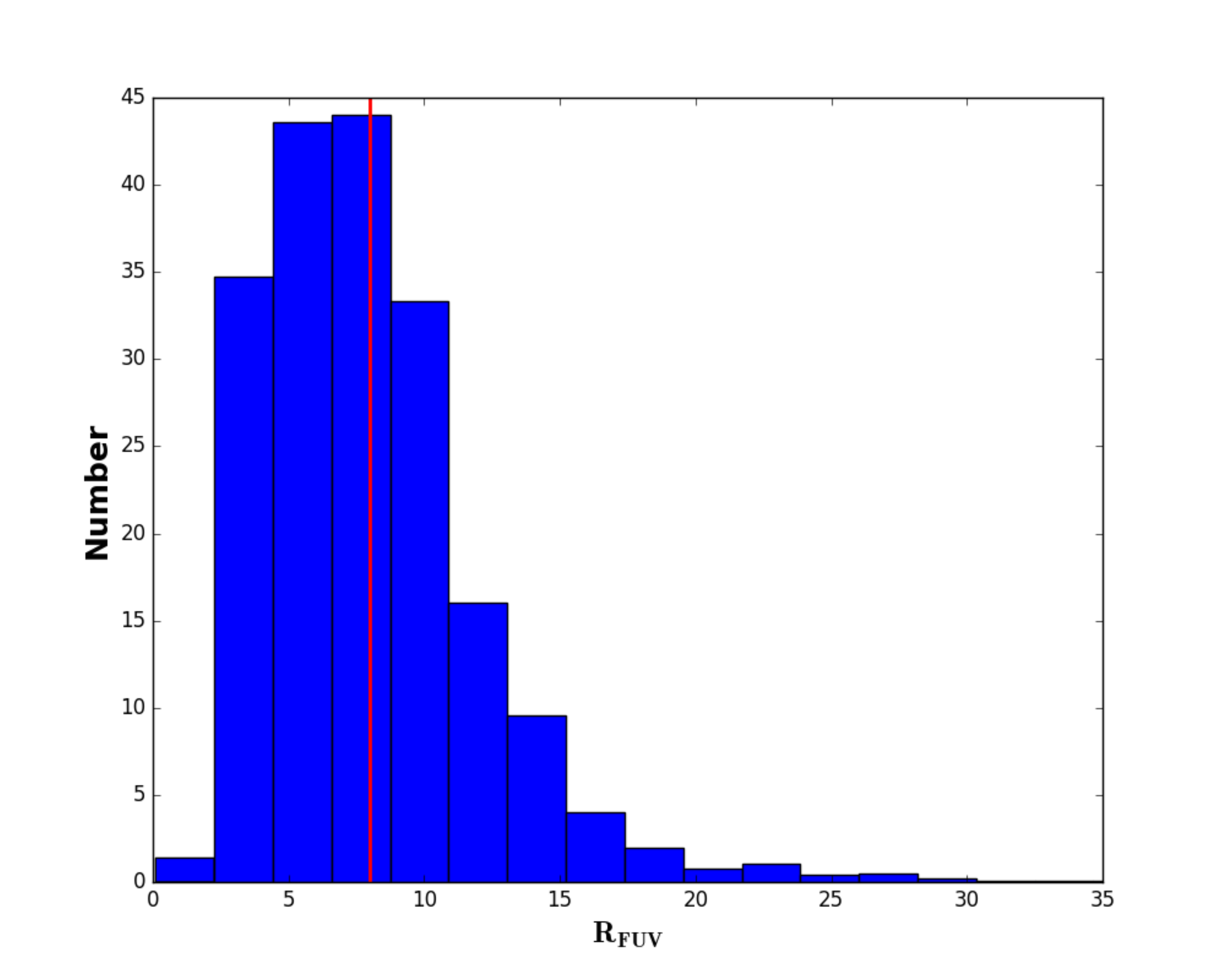}
\includegraphics[width=\columnwidth, bb=38 5 659 539]{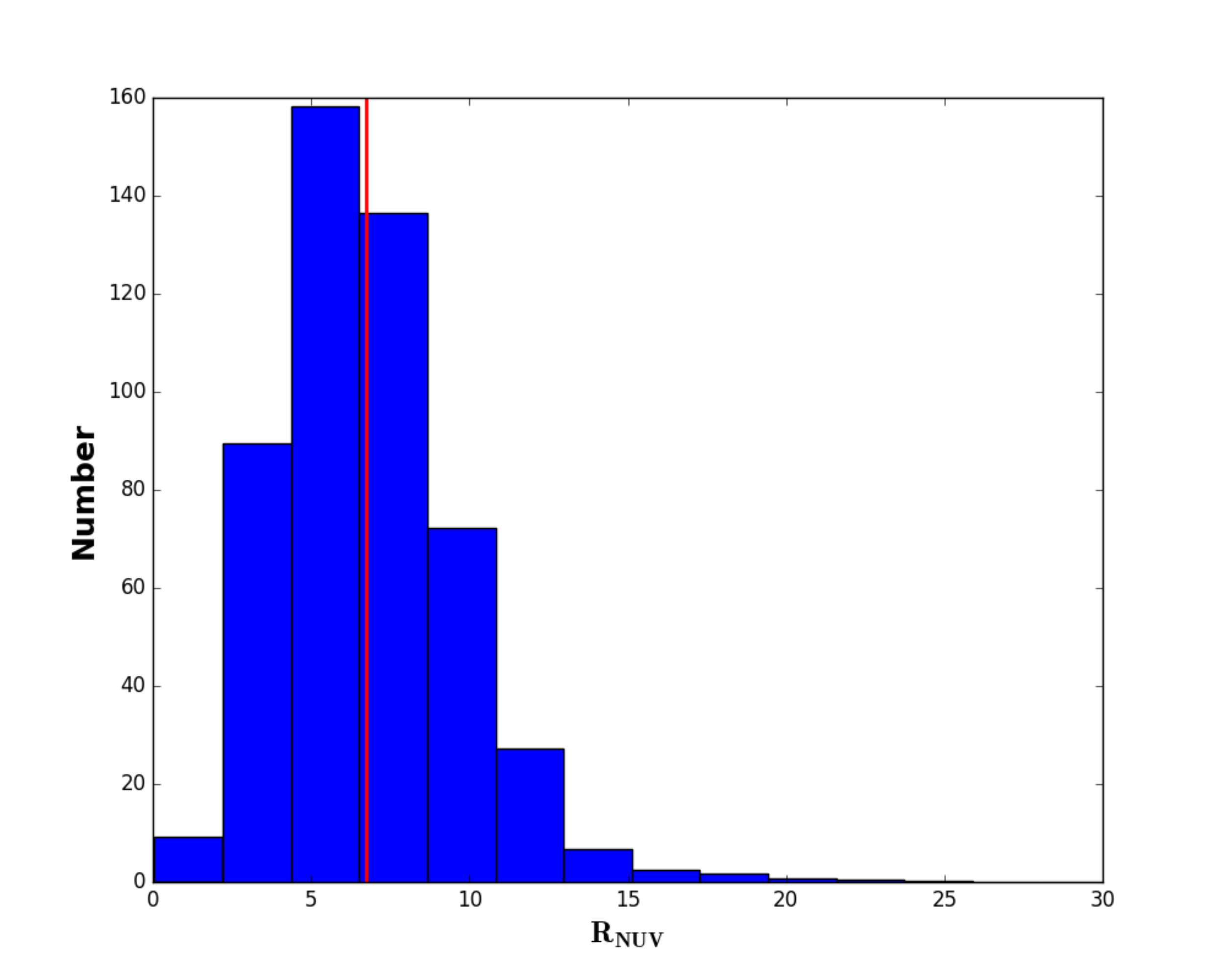}
\caption{Histograms of weighted \textit{R} values for FUV and NUV bands. The solid red line is the weighted average.}
\label{fig:hist}
\end{figure}

Figure \ref{fig:hist}
shows the distribution of the R values in the FUV and NUV filters.
Since the R values for some stars have relatively large uncertainties,
here we plot weighted histograms, where each R value only contributes its
associated error towards the bin count (instead of 1). This figure reveals a relatively 
large spread in R for both filters, with a standard deviation $\sim3$. This spread in R 
values indicates that we cannot characterize the interstellar extinction by a universal
reddening law for all lines of sight within the SDSS footprint. However, our best
estimate, the weighted mean values, are $R_{\rm FUV}=8.01 \pm 0.07$ and $R_{\rm NUV}=6.79 \pm 0.04$.

\begin{figure}
\includegraphics[width=\columnwidth, bb=38 5 659 539]{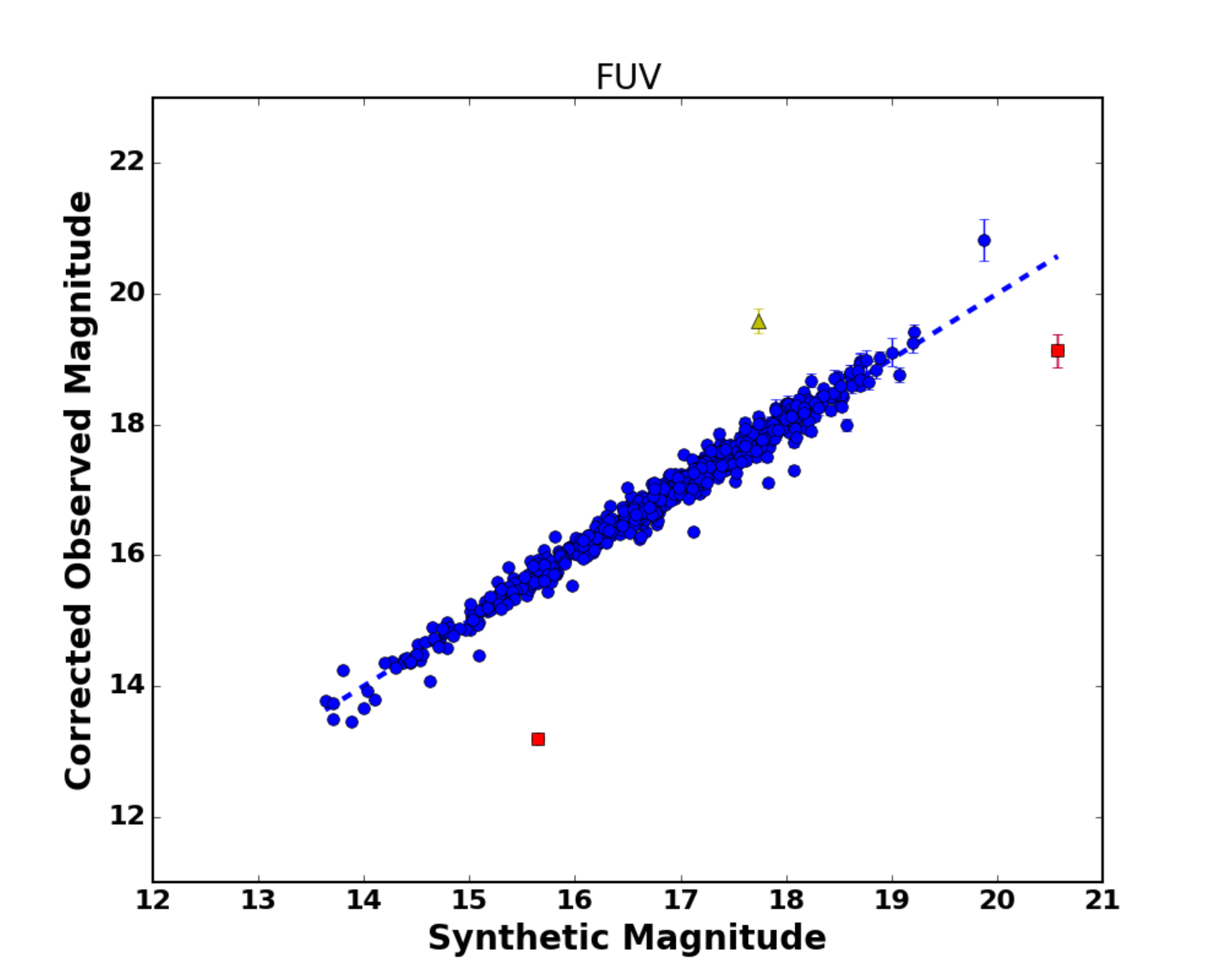}
\includegraphics[width=\columnwidth, bb=38 5 659 539]{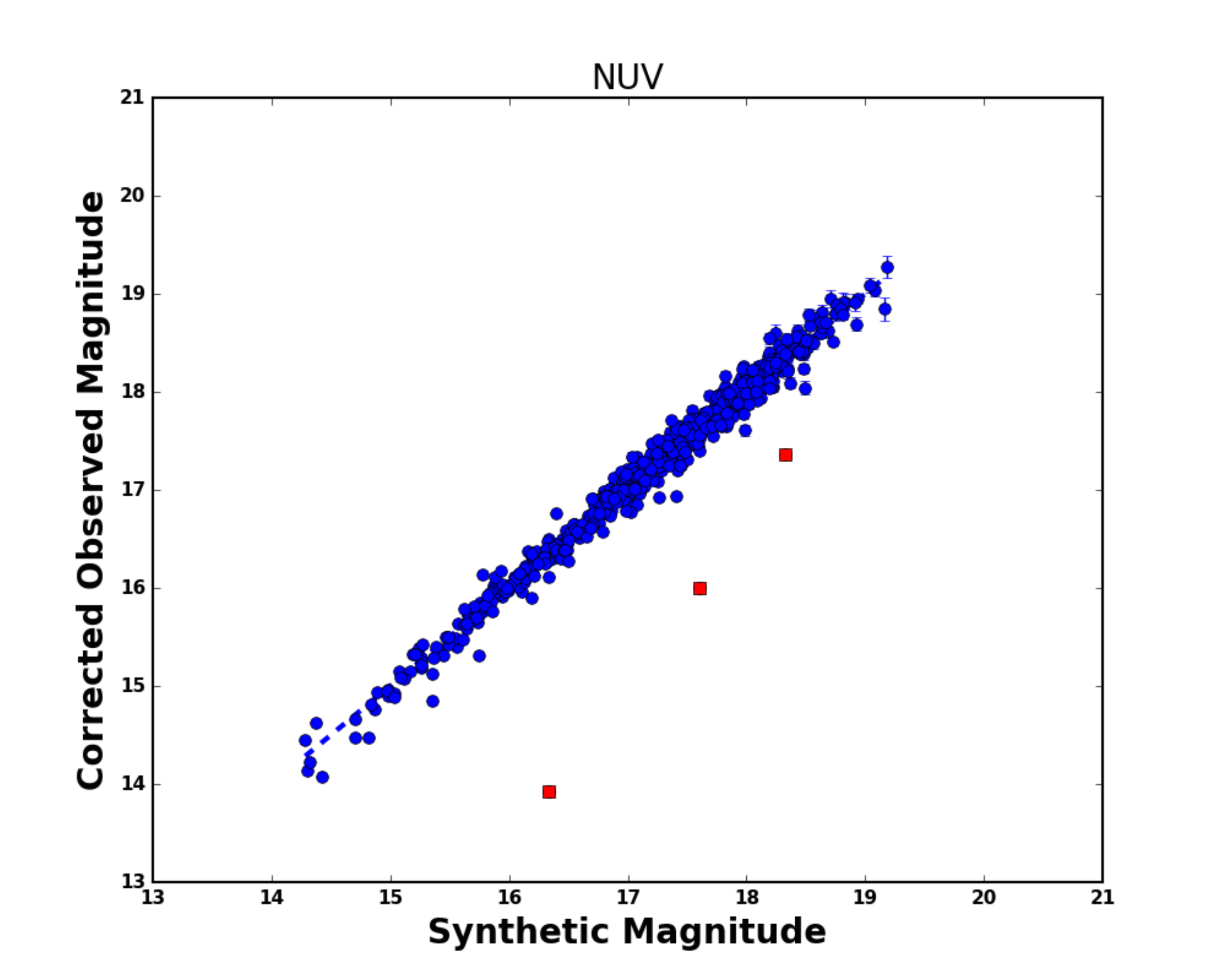}
\caption{Extinction corrected versus model magnitudes for stars beyond
250 pc in the FUV and NUV bands. The blue dashed
line is the one-to-one correlation and 4$\sigma$ outliers are plotted
as red squares. The candidate double degenerate J211607.27+004503.17 is plotted as a yellow triangle.}
\label{fig:with}
\end{figure}

\citet{Bianchi2011} estimated {\em GALEX} extinction coefficients using 
progressively reddened models for stars with $T_{\rm eff}=15,000 - 30,000$ K. 
Since {\em GALEX} NUV band includes the strong broad
absorption feature at 2175 \AA, they predict an overall absorption that is similar
in both the FUV and NUV bands. For Milky Way type dust, they predict
$R_{\rm FUV} \approx R_{\rm NUV} \approx 8.0$. However, for UV-steep extinction curves
like those of the Large Magellanic Cloud (LMC) and the SMC, the increase in absorption
is larger in the FUV and the 2175 \AA\ bump is less pronounced, resulting in estimates
of $R_{\rm FUV}=8.6-12.7$ and $R_{\rm NUV}=7.0-8.1$. Hence, some of the scatter 
seen in Figure \ref{fig:hist} can be explained by the differences
in extinction curves along different line of sights as sampled by our targets. 

Empirical constraints on {\em GALEX}
extinction by \citet{Yuan2013} agree relatively well in the NUV but they differ
significantly in the FUV. \citet{Yuan2013} measure $R_{\rm NUV}= 7.24 \pm 0.08$ or
$7.06 \pm 0.22$ and $R_{\rm FUV}= 4.89 \pm 0.60$ or $4.37 \pm 0.54$. Our FUV extinction
coefficient is significantly larger than the \citet{Yuan2013} estimate and in good
agreement with the \citet{Bianchi2011} estimate. Given the simplicity of DA white dwarf
photospheres, white dwarfs are excellent spectrophotometric standard stars and our 
empirical results are significantly more precise than
previous FUV and NUV extinction coefficient measurements.

Figure \ref{fig:with} shows a comparison
between the observed FUV/NUV magnitudes corrected for non-linearity and extinction
and synthetic magnitudes for the $d>250$ pc sample using our best-estimates of 
$R_{\rm FUV}=8.01 \pm 0.07$ and $R_{\rm NUV}=6.79 \pm 0.04$. These R values provide
 excellent corrections for our dataset, as the majority
of the objects fall on or near the one-to-one line (shown as a blue dashed line).
The red squares mark the $4\sigma$ outliers from the one-to-one line. 
The yellow square marks J211607.27+004503.17, a previously known candidate binary system \citep{Baxter2014}.
We further discuss the unknown outliers in Section \ref{sec:out}.

\section{Outliers}
\label{sec:out}

Here we revisit the 3 and 4$\sigma$ outliers identified in the 100 and 250 pc samples in 
Sections \ref{sec:lin} and \ref{sec:R}, respectively. One possible cause of a significant
difference between the observed and model FUV and NUV magnitudes is the presence of 
an unseen companion. If two stars are sufficiently close together to be unresolved 
in both \textit{GALEX} and the SDSS observations, one could still identify the
binary nature of the system through UV-excess, like the double white dwarf
SDSS J125733.63+542850.5 \citep{Badenes2009,Kulkarni2010,Marsh2011,Bours2015}. Note that
this method only works for systems 
where there is a significant temperature difference between the two white dwarfs.

\begin{table*}
        \centering \caption{Previously known binary or variable white dwarfs that are identifed as outliers in this work.} \label{tab:known}
        \begin{tabular}{ccc}
                \hline Object & Type & Source \\ \hline 
                WD0037-006 & Double-lined double degenerate & \citet{Koester2009}\\
		WD0232+035 & DA+dM & \citet{Kawka2008}\\
                WD0901+140 & Visual double degenerate & \citet{Farihi2005}\\
		WD1019+462 & WD+dM & \citet{Reid1996}\\
                WD1022+050 & Double degenerate & \citet{Bragaglia1995}\\ 
		WD1258+013 & ZZ Ceti & \citet{Bergeron2004}\\
		J211607.27+004503.17 & Double degenerate candidate & \citet{Baxter2014}\\ \hline
        \end{tabular}
\end{table*}

\begin{figure}
\includegraphics[width=\columnwidth]{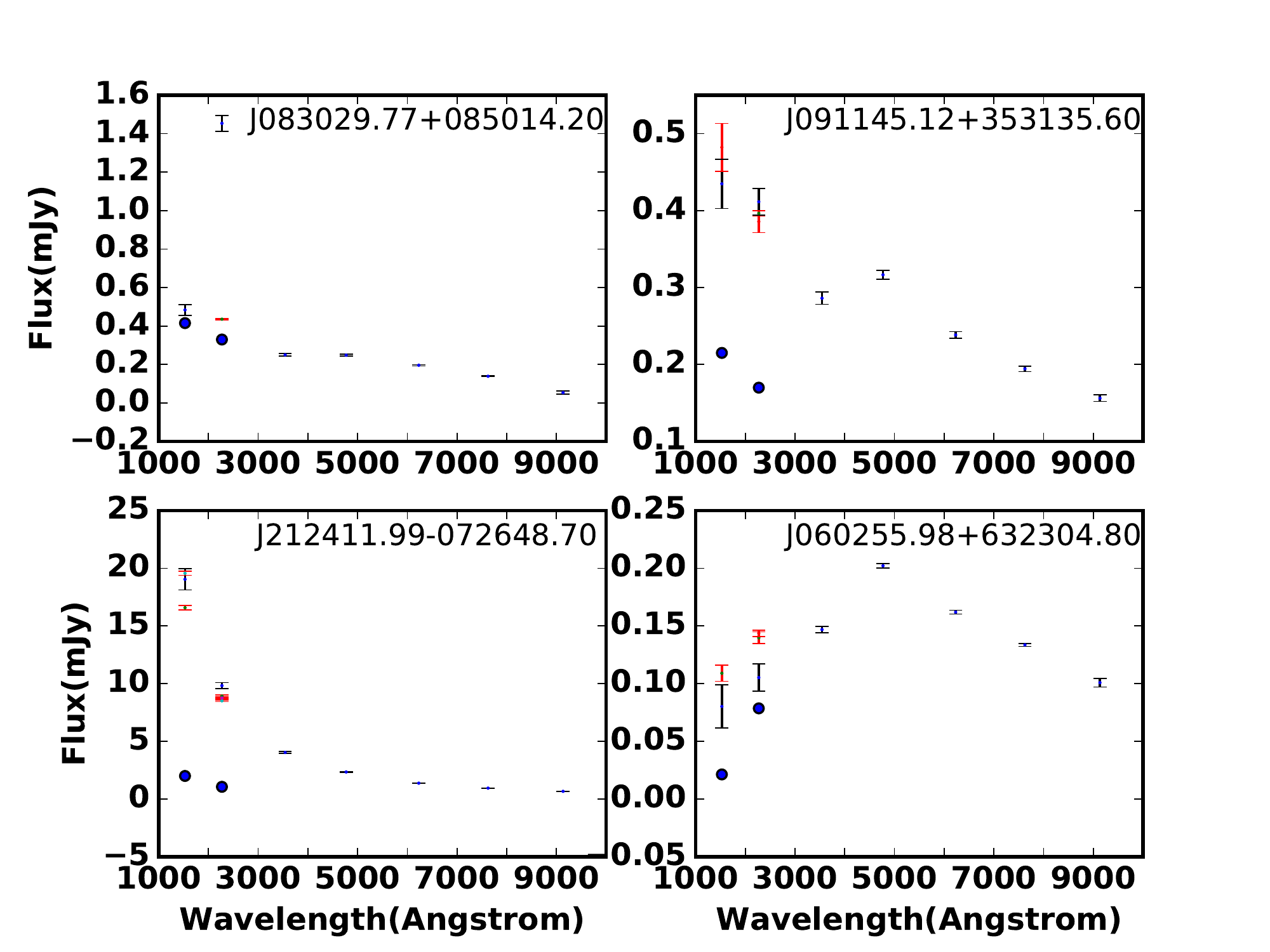}
\caption{SEDs of four newly identified candidate UV-excess white dwarfs. Black errorbars represent fluxes from SDSS and GUVcat. Red errorbars represent fluxes from all other GALEX observations. Blue dots represent model fluxes. Note that two of these outliers, J0830+0850 and J2124-0726, likely
have contaminated photometry or inaccurate photometric solutions.}
\label{fig:out}
\end{figure}

Out of our twelve total outliers, seven are previously known systems. Although WD0901+140 is a visual binary \citep{Farihi2005},
it was not resolved in \textit{GALEX}. Of the five remaining outliers, the photometry of WD0846+335
is likely contaminated by a nearby background galaxy. In Figure \ref{fig:out}, we plot the SEDs
of the remaining four outliers. We plotted the SDSS and GUVcat fluxes as blue errorbars. Each of these
objects has UV observations from other \textit{GALEX} surveys. These fluxes are represented by the red
errorbars. Model fluxes are represented by blue dots. J083029.77+085014.20 is sufficiently near to
a bright star that its photometry was contaminated in the shallow AIS survey. Deeper surveys removed
this contamination, as can be seen in Figure \ref{fig:out}. J212411.99-072648.70 has a spectroscopic $T_{\rm eff}$
of 76,364 K from \citet{Kleinman2013}, well above 35,000 K, the $T_{\rm eff}$ above which the photometric
technique becomes less reliable \citep{Genest-Beaulieu2019}. It is likely that the photometric $T_{\rm eff}$
is off, leading to the apparent UV excess. J060255.98+632304.80 has a $T_{\rm eff}$ of 11,078 K and a $\log{g}$
of 7.7, placing this star within the ZZ Ceti instability strip (10,500-13,000K $T_{\rm eff}$). The UV excess
is greater in other {GALEX} surveys. ZZ Ceti pulsations are stronger in the UV, so the UV excess could
be due to pulsations as with WD1258+013. Further observations are needed to confirm that J060255.98+632304.80
is a ZZ Ceti. There are no obvious explainations for the UV excess of
J091145.12+353135.60. Follow-up UV spectroscopy or radial velocity observations would be helpful in understanding the nature of this object.

\section{Conclusions} 
\label{sec:con} 

We examine a sample of 1837 DA white dwarfs that
were observed by both SDSS and \textit{GALEX}. By combining our SDSS
sample within 100 pc and the bright white dwarf samples form \citet{Holberg2014} and \citet{Gianninas2011},
we determine an improved linearity correction to the
\textit{GALEX} data. We determine that our linearity corrections are only
necessary for objects brighter than 15.95 mag and 16.95 mag for the FUV and 
NUV bands, respectively. We present new extinction
coefficients for the \textit{GALEX} bands: $R_{\rm FUV}=8.01 \pm 0.07$ and $R_{\rm NUV}=6.79 \pm 0.04$.
These white dwarfs currently provide the best constraints on the linearity corrections 
and extinction coefficients for {\em GALEX} data.

Here we present one application of our newly derived
\textit{R} values for identifying unusual white dwarfs.
We identify seven previously known objects: three double degenerates (WD0037-006, WD0901+140, and WD1022+050),
two white dwarf+main sequence binaries (WD1019+462 and WD0232+035), one ZZ Ceti (WD1258+013), and one
double degenerate candidate (J211607.27+004503.17) as outliers.
We find one previously unknown 3$\sigma$ outlier and four previously unknown 4$\sigma$ outliers.
The UV-excesses of three of these objects (WD0846+335, J083029.77+085017.20,
and J212411.99-072648.70) can be explained by contaminating
background sources or inaccurate photometric solutions. Two outliers, J091145.12+353135.60 and J060255.98+632304.80,
require follow-up spectroscopy to verify their natures. In the future, we will use our linearity corrections  
and our newly derived extinction coefficients to study the remainder
of our SDSS sample and identify unusual objects.

\section*{Acknowledgements}

This work is supported by the NASA Astrophysical Data Analysis Program
grant NNX14AF65G.
This work is supported in part by the NSERC Canada and by the Fund FRQ-NT (Qu\'ebec).
This work has made use of data from the European Space Agency (ESA) mission
{\it Gaia} (\url{https://www.cosmos.esa.int/gaia}), processed by the {\it Gaia}
Data Processing and Analysis Consortium (DPAC,
\url{https://www.cosmos.esa.int/web/gaia/dpac/consortium}). Funding for the DPAC
has been provided by national institutions, in particular the institutions
participating in the {\it Gaia} Multilateral Agreement.
We thank our referee, Jay Holberg, for his useful suggestions.

\bibliography{master}
\bsp
\label{lastpage}

\end{document}